\documentclass{emulateapj}

\begin{document}
\slugcomment{Submitted to ApJ Letters}

\title{ A Successful Targeted Search for Hypervelocity Stars }

\author{Warren R.\ Brown\altaffilmark{1},
	Margaret J.\ Geller,
	Scott J.\ Kenyon,
	Michael J.\ Kurtz}

\affil{Smithsonian Astrophysical Observatory, 60 Garden St, Cambridge, MA 02138}
\email{wbrown@cfa.harvard.edu}

\altaffiltext{1}{Clay Fellow, Harvard-Smithsonian Center for Astrophysics}

\shorttitle{ Hypervelocity Stars }
\shortauthors{Brown et al.}

\begin{abstract}

	Hypervelocity stars (HVSs) travel with velocities so extreme that
dynamical ejection from a massive black hole is their only suggested origin.
Following the discovery of the first HVS by Brown and collaborators, we have
undertaken a dedicated survey for more HVSs in the Galactic halo and present
here the resulting discovery of two new HVSs:  SDSS J091301.0+305120 and SDSS
J091759.5+672238, traveling with Galactic rest-frame velocities at least
$+558\pm12$ and $+638\pm12$ km s$^{-1}$, respectively.  Assuming the HVSs are
B8 main sequence stars, they are at distances $\sim$75 and $\sim$55 kpc,
respectively, and have travel times from the Galactic Center consistent with
their lifetimes.  The existence of two B8 HVSs in our 1900 deg$^2$ survey,
combined with the Yu \& Tremaine HVS rate estimates, is consistent with HVSs
drawn from a standard initial mass function but inconsistent with HVS drawn
from a truncated mass function like the one in the top-heavy Arches cluster.  
The travel times of the five currently known HVSs provide no evidence for a
burst of HVSs from a major in-fall event at the Galactic Center in the last
$\sim$160 Myr.

\end{abstract}


\section{INTRODUCTION}

	All galaxies with bulges probably host massive black holes (MBHs)  in
their centers.  \citet{hills88} first showed that a three-body interaction
involving a MBH and a stellar binary can eject one member of the binary with
$>$1,000 km s$^{-1}$ velocity.  Hills called stars ejected with $>$1,000 km
s$^{-1}$ velocities ``hypervelocity stars.'' Hypervelocity stars (HVSs) are
thus a natural consequence of the presence of a massive black hole in a dense
stellar environment.

	\citet{brown05} reported the first discovery of a HVS:  a $g'=19.8$
late-B type star, $\sim$110 kpc distant in the Galactic halo, traveling with a
Galactic rest-frame velocity of at least $+709\pm12$ km s$^{-1}$ (heliocentric
radial velocity +853 km s$^{-1}$).  Photometric follow-up revealed that the
object is a slowly pulsating B main sequence star \citep{fuentes06}.  Only
interaction with a MBH can plausibly accelerate a 3 $M_{\sun}$ main sequence B
star to such an extreme velocity.

	Our HVS discovery inspired a wealth of work from both observers and
theorists.  \citet{edelmann05} report a 8 $M_{\sun}$ main sequence B star,
$\sim$60 kpc distant, traveling with a Galactic rest-frame velocity of at least
$+548$ km s$^{-1}$ that may be a HVS ejected from the LMC.  \citet{hirsch05}
report a helium-rich subluminous O star, $\sim$20 kpc distant, traveling with a
Galactic rest-frame velocity of at least $+717$ km s$^{-1}$ that is probably a
HVS ejected from the Galactic Center.  \citet{holley06} suggest that outliers
in the velocity distribution of intracluster planetary nebulae around M87 may
be HVSs.  In this paper, we report the discovery of two more HVSs from our
ongoing HVS survey.

	With HVSs now an observed class of objects, it is important to define
true HVSs.  Run-away B stars located many kpc above the Galactic plane have
long been known, but their velocities are typically $\lesssim200$ km s$^{-1}$
and they are very probably bound to the Galaxy.  HVSs, on the other hand, are
unbound.  More importantly, the classical supernova ejection \citep{blaauw61}
and dynamical ejection \citep{poveda67} mechanisms that explain run-away B
stars cannot produce ejection velocities which exceed $\sim$300 km s$^{-1}$
\citep{leonard93, gualandris05}.  Thus we define a HVS as an unbound star with
an extreme velocity that can be explained so far only by dynamical ejection
associated with a MBH.

	HVSs are important tools for understanding the nature and environs
of MBHs:
	\citet{holley06} predict that a thin torus of ejected HVSs is the
signature of two MBHs forming a tight binary.
	\citet{ginsburg06} suggest that the stars on highly eccentric orbits
around SgrA$^{*}$ may be the former companion stars to HVSs ejected by the MBH.
	\citet{levin05} shows that an intermediate mass black hole (IMBH) on a
circular in-spiral into the Galactic Center produces an isotropic burst of
HVSs; an IMBH on an eccentric in-spiral produces broad jets of HVSs.
	\citet{gualandris05} find that HVSs produced from stellar binary
encounters with single MBHs have higher ejection velocities than HVSs from
binary MBHs.
	\citet{gnedin05} show that the distance and full space motion
of HVSs can provide significant constraints on the shape and orientation of
the Galactic dark matter halo.
	\citet{yu03} expand Hill's original analysis and show that single star
encounters with binary MBHs produce $\sim$10 times more HVSs than stellar
binary encounters with single MBHs.

	Our paper is organized as follows.  In \S 2 we describe our survey target
selection and observations.  In \S 3 we present the new HVSs.  We conclude in \S 4
by discussing what the observed set of HVSs implies about their origin and the
nature of the Galactic Center.

\section{DATA}

\subsection{Target Selection}

	HVSs ought to be rare:  \citet{yu03} predict there should be
$\sim$10$^3$ HVSs in the entire Galaxy.  Thus, in any search for HVSs, survey
volume is important.  Solar neighborhood surveys have not discovered HVSs
because, even if they were perfectly complete to a depth of $d=1$ kpc, there is
a $\sim$0.1\% chance of finding a HVS in such a small volume.  Finding a new
HVS among the Galaxy's $\sim$10$^{11}$ stars also requires selection of targets
with a high probability of being HVSs.  Our observational strategy is two-fold.  
Because the density of stars in the Galactic halo drops off as approximately
$r^{-3}$, and the density of HVSs drops off as $r^{-2}$ (if they are produced
at a constant rate), we target distant stars where the contrast between the
density of HVSs and indigenous stars is as large as possible.  Secondly, the
stellar halo contains mostly old, late-type stars.  Thus we target faint B-type
stars, stars with lifetimes consistent with travel times from the Galactic
center but which are not a normally expected stellar halo population.  This
strategy makes sense because 90\% of the $K<16$ stars in the central
$0.5\arcsec$ of the Galactic Center are in fact normal main sequence B stars
\citep{eisenhauer05}.

\begin{figure}
 \includegraphics[width=3.25in]{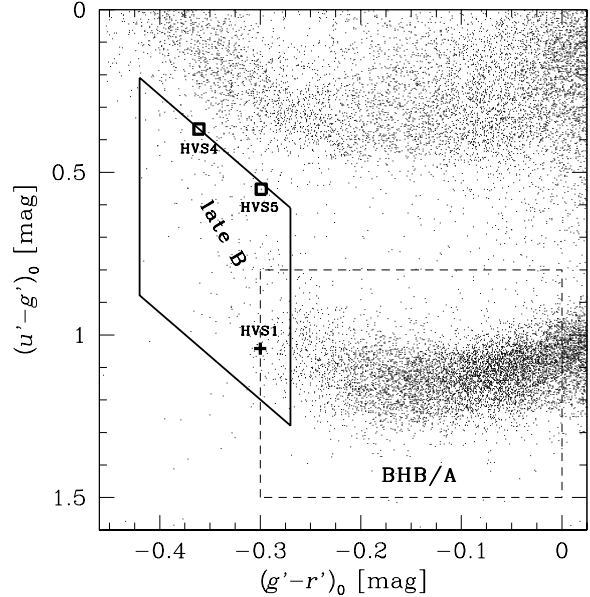}
 \figcaption{ \label{fig:ugr}
	Color-color diagram showing our target selection, illustrated with
every star in the SDSS DR4 $17.5<g'_0<18.5$.  For reference, BHB/A stars are
located in the dashed box \citep{yanny00}.  Candidate late B-type stars are
seen extending up the stellar sequence towards the ensemble of white dwarfs,
and are selected by the solid parallelogram. The original HVS1 \citep{brown05}
is marked by plus sign, and the new HVSs are marked by open squares.}
\end{figure}

	We use SDSS photometry to select candidate B stars by color.  To
illustrate the color selection, Fig.\ \ref{fig:ugr} shows a color-color diagram
of every star with $17.5<g'_0<18.5$ and B- and A-type colors in the SDSS DR4
\citep{adelman06}.  We use de-reddened colors computed from $E(\bv)$ values
obtained from \citet{schlegel98}.  The dashed box indicates the selection
region used by \citet{yanny00} to identify BHB candidates.  Interestingly,
there is a faint group of stars with late B-type colors extending up the
stellar sequence towards the ensemble of white dwarfs with
$(u'-g')_0\lesssim0.5$.  We select candidate B stars inside the solid
parallelogram defined by: $-0.42<(g'-r')_0<-0.27$ and $2.67(g'-r')_0 + 1.33 <
(u'-g')_0 < 2.67(g'-r')_0 + 2.0$.

	We observed a complete sample of 79 candidate B stars in the 1900
deg$^2$ region of the SDSS DR4 bounded by $7^h40^m < {\rm RA} < 10^h50^m$ and
Dec $>15\arcdeg$.  Figure \ref{fig:sky} displays the locations of the objects
on the sky.  The \citet{hirsch05} HVS is located in this region on the sky, but
it is not a part of our survey because its photometric colors lie far outside
of our selection box.  Our sample of candidate B stars is 100\% complete in the
magnitude range $17.0<g'_0<19.5$.

\subsection{Spectroscopic Observations}

	Observations were obtained 2005 December 3-5 with the Blue Channel
spectrograph on the 6.5m MMT telescope.  The spectrograph was operated with the
832 line/mm grating in second order, providing 1.2 \AA\ spectral resolution
and wavelength coverage 3660 \AA\ to 4500 \AA.  Exposure times ranged from 5 to
30 minutes and were chosen to yield $S/N=15$ in the continuum at 4000 \AA.  
Comparison lamp exposures were after obtained after every exposure.

	Radial velocities were measured using the cross-correlation package
RVSAO \citep{kurtz98}.  The average uncertainty is $\pm12$ km s$^{-1}$.  We
correct the heliocentric velocities to the local standard of rest
\citep{hogg05} and remove the 220 km s$^{-1}$ solar reflex motion.  Thus all
velocities reported here are in the Galactic rest frame, indicated $v_{RF}$.  
The velocities of the candidate B stars are indicated by color in Fig.\
\ref{fig:sky}.

\subsection{Selection Efficiency and Unusual Objects}

	Our sample of 79 targets is composed of 61 late B-type stars, 16 faint
DA white dwarfs, and 2 low-$z$ galaxies.  We derive spectral types of the late
B-type stars based on line indices described in \citet{brown03}; the types
range from B6 to A1.  Thus our target selection is 77\% efficient for selecting
stars of late B spectral type.  Changing the color selection edge from
$(g'-r')_0=-0.42$ to $-0.38$ would eliminate more than half of the white dwarfs
and one of the galaxies, and would increase the target selection efficiency for
late B-type stars to 90\%.  We plan to publish the faint white dwarfs and other
unusual objects in a future paper describing the HVS survey as a whole.

\section{HYPERVELOCITY STARS}

\begin{figure}
 \includegraphics[width=3.25in]{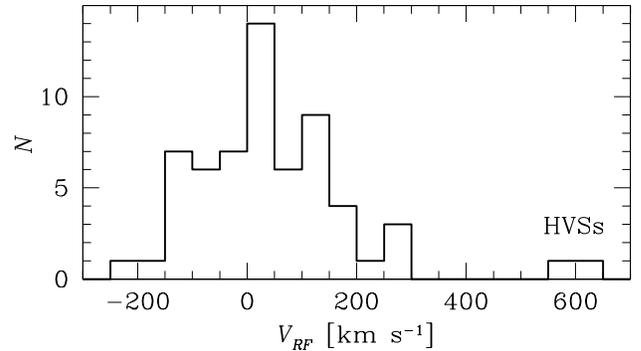}
 \figcaption{ \label{fig:velh}
	Velocity histogram of the late B-type stars.  Two new HVSs are
identified moving away with Galactic rest-frame velocities at least $+558$ and
$+638$ km s$^{-1}$.}
\end{figure}

	Our targeted search for HVSs uncovered two new HVSs, SDSS
J091301.0+305120 (hereafter HVS4) and SDSS J091759.5+672238 (hereafter HVS5).  
Figure \ref{fig:velh} plots a histogram of Galactic rest-frame radial velocity
for the 61 late B-type stars in our sample.  Ignoring the HVSs, our sample has
a velocity dispersion of $\pm114$ km s$^{-1}$ consistent with a Galactic halo
population.  HVS4 and HVS5 are traveling with $v_{RF}=+558\pm12$ and
$+638\pm12$ km s$^{-1}$ (heliocentric radial velocities $+603$ and $+543$ km
s$^{-1}$), respectively, and are 5-$\sigma$ outliers from this distribution.  
The escape velocity of the Galaxy is approximately 300 km s$^{-1}$ at 50 kpc
\citep{wilkinson99}; thus HVS4 and HVS5 are clearly unbound to the Galaxy.

\begin{figure}
 \includegraphics[width=3.25in]{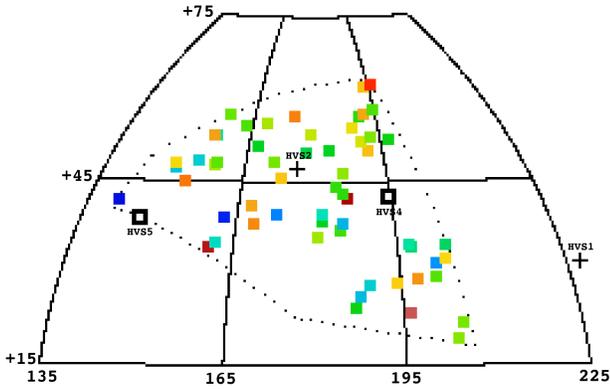}
 \figcaption{ \label{fig:sky}
	Aitoff sky map, in Galactic coordinates, showing the complete sample of
61 late B-type stars.  Dotted lines indicate the 1900 deg$^2$ region of the
SDSS DR4 we observed.  Radial velocities, in the Galactic rest frame, are
indicated by the color of the solid squares:  purple is -300, green is 0, and
red is +300 km s$^{-1}$.  The new HVSs are completely off this color scale and
are marked by open squares; HVS1 \citep{brown05} and HVS2 \citep{hirsch05} are 
marked by plus signs. }
\end{figure}

	The new HVSs are not physically associated with any other Local Group
galaxy.  HVS4 and HVS5 are located at $(l,b)=(194.8^{\arcdeg},42.6^{\arcdeg})$
and $(146.3^{\arcdeg},38.7^{\arcdeg})$, respectively (see Fig.\ \ref{fig:sky}).  
HVS4 is separated from Leo A by 10$^{\arcdeg}$ on the sky, but the galaxy is in
the distant background 800 kpc away \citep{dolphin02}.  Even if HVS4 were a
$M_V=-6$ supergiant at the distance of Leo A, the star's radial velocity
differs from Leo A by 575 km s$^{-1}$.  The closest galaxy to HVS5 is the Ursa
Minor dwarf 31$^{\arcdeg}$ away, yet the velocity difference is even more
extreme at 730 km s$^{-1}$.

	Both HVS4 and HVS5 have spectral types consistent with B8, however our
low-resolution spectra do not allow us to determine exact stellar parameters.  
Stars of this spectral type are likely blue horizontal branch (BHB) stars or
main sequence B stars/blue stragglers.  Two of the three previously reported
HVSs are main sequence B stars \citep{fuentes06,edelmann05}.  
	We note that the Balmer line widths of HVS4 and HVS5 are too broad to
be consistent with those of B I supergiants.
	If we assume the HVSs are BHB stars rather than B stars, their blue
colors means they are hot, extreme BHB stars and thus they are intrinsically
very faint.  The $M_V(BHB)$ relation of \citet{clewley02} yields
$M_V(BHB)\simeq+2.0$ and $+1.5$ and heliocentric distance estimates
$d_{BHB}\simeq$20 and 18 kpc for HVS4 and HVS5, respectively.  
	In the BHB interpretation, the volume we effectively survey is much
smaller than in the B star interpretation. Thus the BHB interpretation requires
more than an order of magnitude larger production rate for HVSs. Because two of
the previous HVS are B stars and because the B star interpretation implies a
lower production rate probably consistent with \citet{yu03}, we assume that
HVS4 and HVS are B stars for the purpose of discussion.
	The ultimate discriminant will come from higher resolution, higher
signal-to-noise spectroscopy.

	We estimate the luminosity of a B8 star from the \citet{schaller92}
stellar evolution tracks for a 4 $M_{\sun}$ star with $Z=0.02$.  Such a star
spends 160 Myr on the main sequence and produces 400 $L_{\sun}$ at
$T_{eff}\sim13,000$.  This luminosity corresponds to $M_V(B8)\simeq-0.9$ (using
bolometric correction $-0.80$ \citep{kenyon95}) and heliocentric distances
$d_{B8}\simeq75$ and 55 kpc for HVS4 and HVS5, respectively.  Assuming the Sun
is 8 kpc distant from the Galactic center, the Galacto-centric distances of
HVS4 and HVS5 are thus $r\simeq80$ and 60 kpc, respectively.

	If HVS4 and HVS5 originate from the Galactic Center, the travel times
to their present locations are approximately 140 and 90 Myr, respectively,
consistent with the 160 Myr main sequence lifetime of a 4 $M_{\sun}$ star.  If
the HVSs are BHBs or blue stragglers, their lifetimes are considerably longer
and the travel time constraint is relaxed.

	Our radial velocities provide only a {\it lower} limit to the HVS's
true space velocities.  HVS4 and HVS5 have bright apparent magnitudes
($g'=18.40$ and 17.93, respectively) and are present in the USNOB1
\citep{monet03} and GSC2.3 (B.\ McLean, 2006 private communication) catalogs.  
However, the HVSs are listed with no measurable proper motions, consistent with
the estimated distances.

	Table \ref{tab:hvs} summarizes the properties of all five known HVSs.  
The columns include HVS number, Galactic coordinates $(l,b)$, apparent
magnitude $g'$, minimum Galactic rest-frame velocity $v_{RF}$ (not a full space
velocity), heliocentric distance estimate $d$, travel time estimate from the
Galactic Center $t_{GC}$, and catalog identification.  We note that the B9
$M_V$ estimate was incorrect in \citet{brown05} and should be $M_V(B9)=-0.3$.  
Thus the correct distance and travel time estimates to HVS1 are $d\sim$110 kpc
and $t_{GC}\sim160$ Myr, as indicated.

\section{DISCUSSION}

	The very existence of the HVSs places interesting limits on the stellar
mass function of HVSs, the origin of massive stars in the Galactic Center, and
the history of stellar interactions with the MBH.  Interestingly, all five
known HVSs are moving with positive radial velocity, consistent with the
picture of a Galactic Center origin.

	In principle, we can constrain the stellar population from which the
HVSs originate by combining predictions of HVS rates with the results of our
survey.  \citet{yu03} predict a HVS rate of $\sim$10$^{-5}$ yr$^{-1}$ ejected
by a single MBH at the Galactic Center.  A HVS moving 600 km s$^{-1}$ travels
120 kpc in 200 Myr, thus the \citet{yu03} rate implies $\sim$2000 HVSs of all
types to a depth of 120 kpc.  Our $g'_0<19.5$ magnitude-limited survey reaches
the same depth $d=120$ kpc for B8 stars and we find two B8 HVSs over our 1900
deg$^2$ region, implying $43\pm31$ B8 HVSs over the entire sky.  We ask whether
this number of B stars is consistent with the number expected from a standard
initial mass function (IMF).  A Salpeter IMF \citep{salpeter55}, integrated
over the mass range 0.2-100 $M_{\sun}$ and normalized to 2000 stars, contains
26 stars between 3 and 5 $M_{\sun}$.  A Scalo IMF \citep{scalo86}, similarly
calculated, contains 13 stars between 3 and 5 $M_{\sun}$.  The IMF predictions
are systematically lower than the observed frequency implied by the two B stars
in our survey, but because of small number statistics there is no significant
inconsistency.

	In the Galactic center, there is some indication that the IMF is top
heavy.  For example, \citet{stolte05} study the mass-segregated Arches cluster
and argue for truncation at masses less than 6 $M_{\sun}$.  Integrating the
\citet{stolte05} mass function $\Gamma=-0.86$ over the mass range 3-100
$M_{\sun}$ and normalizing it to 2000 stars results in 750 3-5 $M_{\sun}$
stars, an order of magnitude more late-B HVSs than implied by our
observations.  Thus our observations already indicate that the HVS parent
population is not composed entirely of clusters like Arches or the \citet{yu03} 
HVS rate is an underestimate.

	HVS travel times from the Galactic Center constrain the history of
stellar interactions with the MBH.  Assuming the new HVSs are B stars, travel
time estimates for the known HVSs are spread rather uniformly between 30 and
160 Myr.  Thus there is as yet no evidence for a burst of HVSs from major
in-fall event at the Galactic Center in the last $\sim$160 Myr.  The current
constraints are less clear if HVS4 or HVS5 are not B stars.  If new discoveries
of HVSs continue to show no evidence for coherent bursts of HVSs, theories of
in-falling massive star clusters \citep{gerhard01,kim03} possibly containing an
IMBH \citep{hansen03} to transport B stars near the MBH may not be applicable
to the Galaxy.

	It is interesting that the northern HVSs all have similar
$b\sim40^{\arcdeg}$ (see Fig.\ \ref{fig:sky}).  An in-spiraling binary black
hole produces an unique distribution of HVSs on the sky \citep{levin05,
holley06}.  Detailed theoretical predictions of the distribution of HVSs on the
sky ejected by a single MBH at the Galactic Center would be an important
testbed for future HVS observations.

	HVSs are becoming important tools for understanding MBHs.  Discovering
additional HVSs in a well-defined volume will provide better constraints on the
origin of HVSs and the nature of the Galactic Center.  Proper motion
measurements of HVSs, measured with the {\it Hubble Space Telescope}, the {\it
Global Astrometric Interferometer for Astrophysics}, or the {\it Space
Interferometry Mission}, may provide significant constraints on the shape of
the Galaxy's dark matter potential \citep{gnedin05}.  The distribution of HVSs
in velocity and space may constrain the history of stellar interactions with
the MBH.  Identifying HVSs around other galaxies \citep{holley06} is also an
exciting prospect.  We are continuing our radial velocity survey of every late
B-type star in the SDSS.


\acknowledgements
	We thank J.\ McAfee for his assistance with observations obtained at
the MMT Observatory, a joint facility of the Smithsonian Institution and the
University of Arizona.  This work was supported by W.\ Brown's Clay Fellowship
and the Smithsonian Institution.





\begin{deluxetable}{lccccccl}		
\tablewidth{0pt}
\tablecaption{HYPERVELOCITY STARS\label{tab:hvs}}
\tablecolumns{8}
\tablehead{
	\colhead{ID} & \colhead{$l$} & \colhead{$b$} & \colhead{$g'$} &
	\colhead{$v_{RF}$} & \colhead{$d$} & \colhead{$t_{GC}$} & 
	\colhead{Catalog} \\
	\colhead{} & \colhead{{\small deg}} & \colhead{{\small deg}} & 
	\colhead{{\small mag}} & \colhead{{\small km s$^{-1}$}} & 
	\colhead{{\small kpc}} & \colhead{{\small Myr}} & \colhead{}
}
	\startdata
HVS1 & 227.3 & ~31.3 & 19.8 & +709 & 110 & 160 & SDSS J090745.0+024507$^1$ \\
HVS2 & 176.0 & ~47.1 & 18.8 & +717 & ~19 & ~32 & US 708$^2$ \\
HVS3 & 263.0 & -40.9 & 16.2 & +548 & ~61 & 100 & HE 0437-5439$^3$ \\
HVS4 & 194.8 & ~42.6 & 18.4 & +558 & ~75 & 140 & SDSS J091301.0+305120 \\
HVS5 & 146.3 & ~38.7 & 17.9 & +638 & ~55 & ~90 & SDSS J091759.5+672238 \\
	\enddata 
\tablerefs{ (1) \citet{brown05}; (2) \citet{hirsch05}; (3) \citet{edelmann05} }
 \end{deluxetable}

\end{document}